\documentclass[12pt]{article}
\usepackage{amsfonts}
\usepackage{amssymb}
\usepackage{amsthm}
\newcommand{\maxtensor}{\otimes_{max}}
\newcommand{\mintensor}{\otimes_{min}}

\renewcommand{\hat}{\widehat}

\newcommand{\Hom}{\mbox{Hom}}

\newcommand{\interior}{\mbox{int}}

\newcommand{\ve}[1]{{\bf #1}}

\renewcommand{\phi}{\varphi}
\renewcommand{\H}{{\bf H}}

\newcommand{\tr}{\text{Tr}}

\newcommand{\beq}{\begin{equation}}
\newcommand{\eeq}{\end{equation}}
\newcommand{\beqa}{\begin{eqnarray}}
\newcommand{\eeqa}{\end{eqnarray}}
\def\cl{{\cal L}}

\def\R{{\mathbb R}}

\def\H{{\mathbb H}}

\theoremstyle{plain}
\newtheorem{theorem}{Theorem}
\newtheorem{proposition}[theorem]{Proposition}
\newtheorem{lemma}[theorem]{Lemma}
\newtheorem{corollary}[theorem]{Corollary}

\theoremstyle{definition}
\newtheorem{definition}[theorem]{Definition}

\theoremstyle{remark}
\newtheorem{example}[theorem]{Example}
\newtheorem{examples}[theorem]{Examples}

\def\id{{\rm id}}
\def\tr{\mbox{Tr}}

\newcounter{thingy}
\renewcommand{\thethingy}{\arabic{section}.\arabic{thingy} }

\renewcommand{\section}[1]{ \refstepcounter{section}
\setcounter{thingy}{0}
{ \bf \arabic{section}.~#1 }}{}

\renewcommand{\subsection}[1]{ \refstepcounter{subsection}
\setcounter{thingy}{0}
{ \bf \arabic{section}.\arabic{subsection}~#1 }}{}

\title{Ordered Linear Spaces and Categories as Frameworks for 
Information-Processing Characterizations of Quantum and Classical Theory}

\author{Howard Barnum and Alexander Wilce}

\date{March 2, 2009}

\begin{document}

\maketitle

%\iffalse

\begin{abstract}
The advent of quantum computation and quantum information science
has been accompanied by a revival of the project of characterizing
quantum and classical theory within a setting significantly more
general than both.  Part of the motivation is to obtain a clear
conceptual understanding of the sources of quantum theory's
greater-than-classical power in areas like cryptography and
computation, as well as of the limits it appears to share with
classical theory.
%, such as the apparent inability to solve NP-complete
% problems in polynomial time.  
This line of work suggests
supplementing traditional approaches to the axiomatic
characterization of quantum mechanics within broader classes of
theories, with an approach in which some or all of the axioms
concern the information-processing power of the theory.

In this paper, we review some of our recent results (with
collaborators) on information processing in an ordered linear spaces
framework for probabilistic theories.  These include demonstrations
that many ``inherently quantum" phenomena are in reality quite general
characteristics of {\em non-classical} theories, quantum or otherwise.
As an example, a set of states in such a theory is broadcastable if,
and only if, it is contained in a simplex whose vertices are
cloneable, and therefore distinguishable by a single measurement.  As
another example, information that can be obtained about a system in
this framework without causing disturbance to the system state, must
be inherently classical.  We also review results on teleportation
protocols in the framework, and the fact that any non-classical theory
without entanglement allows exponentially secure bit commitment in
this framework.  Finally, we sketch some ways of formulating our
framework in terms of categories, and in this light consider the
relation of our work to that of Abramsky, Coecke, Selinger, Baez and
others on information processing and other aspects of theories
formulated categorically.

\end{abstract}
%\fi

\newpage
\section{Introduction}

%%FDBNS?

The advent of quantum information theory has been accompanied by a
resurgence of interest in the convex, or ordered linear spaces,
framework for operational theories, as researchers seek to
understand the nature of information processing in increasingly
abstract terms, both in order to illuminate the sources of the
difference between the information processing power of quantum
versus classical theory, and because quantum information has
occasioned renewed interest in foundational aspects of quantum
theory, often with the new twist that axioms or principles
concerning information processing are considered. A representative
(but by no means exhaustive) sample might include the work of Hardy \cite{Hardy2001a,Hardy2001b},
D'Ariano \cite{d'Ariano2006a}, Barrett \cite{Barrett:2005a}, and others. The general drift of this work is to
suggest that it may be possible to characterize quantum mechanics
largely or entirely in terms of informational properties, and especially
its information processing capabilities.  A particularly sharp conjecture
advocated by Brassard \cite{Brassard:2005a} and
Fuchs \cite{Fuchs:2002, Fuchs:2003} suggests that QM may be uniquely characterized by three
information-theoretic constraints: a no-signaling constraint on
compound systems, the impossibility of bit commitment, and the
possibility of tamper-evident key distribution using a small amount
of authenticated public classical communication.
%Clifton, Bub, and
%Halvorson [ref] showed that in $C^*$-algebraic theories with a
%notion of locality that prohibits signaling, no-broadcasting
%($\equiv$ noncommutativity; a proxy for key distribution) and {\em
%strong} no-bit-commitment imply the existence of entangled states,
%by providing a (weak) bit commitment protocol for any such theory
%without entanglement. However, their NBC assumption
%is very strong, requiring that no $\epsilon$-hiding
%protocols exist for any $\epsilon <1$. Moreover, in finite dimensions
%$C^*$-algebras are essentially QM with ``superselection sectors', so the result's
%foundational significance in this setting is limited. Finally, Halvorson has
%since shown that no-broadcasting \alert{{\em alone}} $\implies$
%entanglement, so ....

In order to address such questions, it is necessary to have a clear
picture of the information-theoretic properties of probabilistic
theories more general than quantum mechanics. Working in a
well-established mathematical framework for such theories, in which
compact convex sets serve as state spaces, we and our coauthors (Jonathan
Barrett and Matthew Leifer) have shown that many
``characteristically quantum" phenomena---in particular, many aspects
of entanglement, as well as no-cloning and no-broadcasting 
theorems---are quite generic features of a wide class of non-classical
probabilistic theories in which systems are coupled subject to
no-signaling and local observability requirements. In \cite{BDLT2007a},
Barnum, Dahlsten, Leifer and Toner have shown that the impossibility
of a bit-commitment protocol for non-classical state spaces (in the
same broad framework) implies the existence of entangled states.  
Since most ways of combining nonclassical state spaces yield entanglement,
this is only a weak constraint on theories.  Of course, the impossibility
of bit commitment might imply much more about theories, but we do not
yet have such results.  

This suggests that the impossibility of bit-commitment is in fact not a
particularly strong constraint on probabilistic theories. 
%%AW: How does it suggest this? Presumably the idea is that entanglement is very generic. Suggest we say 
%%instead something like, ``...; but, as entanglement is a generic feature of non-classical theories, this is a weak constraint."
On the other hand, as discussed in \cite{BBLW08}, the existence of a
teleportation protocol {\em is} known to be a strong constraint, moving one
somewhat closer to quantum theory.  Even so, one can construct
many models of teleportation---and even of deterministic
teleportation---that are neither classical nor quantum.  The first
part of this contribution gives a concise overview of this work.

In a different direction, with impetus from linear logic and topological
quantum field theory,
various authors---in particular Abramsky and Coecke
\cite{Abramsky:2004a}, Baez \cite{Baez04a}, and Selinger
\cite{Selinger2004a, Selinger2007a}---have established that many of the most striking
phenomena associated with quantum information processing notably,
various forms of teleportation, as well as restrictions on
cloning---arise much more generally in any compact closed category,
including, for instance, the category of sets and relations. In the
second part of this paper, we make a preliminary attempt to relate
this approach to ours.
%The abstract state spaces that we consider
%naturally form a category; however, this is far from being compact
%closed. For one thing, the dual of an abstract state space is
%usually not, in any natural way, another state space, but a
%different sort of beast altogether. Nor are our categories generally
%monoidal: more typically, they support a profusion of possible
%mechanisms for coupling systems, bounded by a maximal (and maximally
%entangled) tensor product $\maxtensor$, and a minimal (unentangled)
%product $\mintensor$. On the one hand, there are various
%constructions by which one can embed our category of state spaces in
%a larger category of {\em processes} having a better behaved -- in
%particular, monoidal and self-dual -- structure. Moving in the
%opposite direction, one can focus on restricted categories that are,
%in a sense (made precise below) ``closed under teleportation": as it
%happens, the entangled state and effect corresponding to a
%correction-free teleportation protocol are precisely the unit and
%co-unit of a duality.

\section{General Probabilistic Theories}

There is a well-established mathematical framework for generalized
probability theory, based on ordered linear spaces, deriving from the
work of Mackey \cite{Mackey:1963} in the late 1950s and further
developed by Ludwig \cite{Ludwig64a, Ludwig67a, Ludwig83a, Ludwig85a},
Davies and Lewis \cite{Davies70a}, Edwards \cite{Edwards:1970a},
Holevo \cite{Holevo:1983} and others in the 1960s and 70s. This
section gives a whirlwind overview of this framework, focusing on
finite-dimensional systems.
%%[We should perhaps say more about finite-dimensionality somewhere..]

\subsection{Abstract state spaces}

By an {\em abstract state space}, we mean a pair $(A,u_{A})$ where
$A$ is a finite-dimensional ordered real vector space, with closed,
generating positive
cone $A_{+}$, and where $u_{A} : A \rightarrow {\Bbb R}$ is a
distinguished linear functional, called the {\em order unit}, that
is {\em strictly} positive on $A_{+} \setminus \{0\}$. A state is
{\em normalized} iff $u_{A}(\alpha) = 1$.   (Henceforth, when we say 
just ``cone'', we will mean a positive, generating cone.)  We write $\Omega_{A}$ for
the convex set of normalized states in $A_{+}$. By way of
illustration, if $A$ is the space ${\Bbb R}^{X}$ of real-valued
functions on a set $X$, ordered pointwise on $X$, with $u_{A}(f) =
\sum_{x \in X} f(x)$, then $\Omega_{A} = \Delta(X)$, the simplex of
probability weights on $X$. If $A$ is the space ${\cal L}(\H)$ of
hermitian operators on a (finite-dimensional) complex Hilbert space
$\H$, with the usual operator ordering, and if the unit is defined
by $u_{A}(a) = \tr(a)$,
then $\Omega_{A}$ is the set of density operators on $\H$. On any
abstract state space $A$, there is a canonical norm (the {\em base
norm}) such that for $\alpha \in A_{+}$, $\|\alpha\| =
u_{A}(\alpha)$. For ${\Bbb R}^{X}$, this is the $\ell^{1}$-norm;
for ${\cal L}(\H)$, it is the trace norm ($||L|| := \tr \sqrt{L^\dagger L}$).

\subsection{Events and Processes}

Events (e.g., measurement outcomes) associated with an abstract
state space $A$ are represented by {\em effects}, i.e.,  positive
linear functionals $a \in A^{\ast}$, with $0 \leq a \leq u_{A}$ in
the dual ordering. Note that $0$ and $u_{A}$ are, by definition, the
least and greatest effects. If $\alpha$ is a {\em normalized} state
in $A$---that is, if $u_{A}(\alpha) = 1$---then we interpret
$a(\alpha)$ as the {\em probability} that the event represented by
the effect $f$ will occur if measured. Accordingly, a discrete {\em
observable} on $A$ is a list $(a_1,...,a_n)$ of effects with $a_1 +
a_2 + \cdots + a_n = u_{A}$, representing a collection of events --
the possible outcomes of an experiment, for instance--- {\em one}
of which must certainly occur. We represent a physical process with
initial state space $A$ and final state space $B$ by a positive
mapping $\tau : A \rightarrow B$ such that, for all $\alpha \in
A_{+}$, $u_{B}(\tau(\alpha)) \leq u_{A}(\alpha)$---equivalently,
$\tau$ is norm-contractive. We can regard $\|\tau(\alpha)\| =
u_{B}(\tau(\alpha))$ as the probability that the process represented
by $\tau$ takes place in initial state $\alpha$; this {\em event} is
represented by the effect $u_{B} \circ \tau$ on $A$.

It is important to note that, in the framework just outlined, the
state space $A$ and its dual space $A^{\ast}$ have (in general) 
quite different structures: $A$ is a {\em cone-base space}\footnote{Often
termed a {\em base-norm space}, when considered as a 
normed ordered space with  the base norm discussed above.}, i.e., an
ordered space with a preferred base, $\Omega_A$, for $A_+$, while
$A^{\ast}$ is an {\em order-unit} space, i.e., an ordered space with
a preferred {\em element} in the interior of its positive cone. Indeed, the spaces
$A$ and $A^{\ast}$ are generally not even isomorphic as ordered
spaces. Where there exists a linear order-isomorphism (that is, a
positive linear mapping with positive inverse) between $A$ and
$A^{\ast}$, we shall say that $A$ is {\em weakly self-dual}. Where
this isomorphism induces an inner product on $A$ such that $A_{+} =
\{ b \in A | \langle b, a \rangle \geq 0 \ \forall a \in A_{+}\}$,
we say that $A$ is {\em self-dual}. Finite dimensional quantum and
classical state spaces are self-dual in this sense. A celebrated
theorem of Koecher \cite{Koecher:58} and of Vinberg \cite{Vinberg60} tells us that
if $A$ is an irreducible, finite-dimensional self-dual state space,
and if the group of affine automorphisms of $A_{+}$ acts
transitively on the interior of $A_{+}$, then the space $\Omega_{A}$
of normalized states is affinely isomorphic to the set of density
operators on an $n$-dimensional real, complex or quaternionic
Hilbert space, or to a ball, or to the set of $3 \times 3$ trace-one
positive semidefinite matrices over the octonions.

\subsection{Information and disturbance}

With Barrett and Leifer, we have shown (as described in
\cite{Barrett:2005a}) that in nonclassical theories, the only information
that can be obtained about the state without disturbing it is
inherently classical information---information about which of a set
of irreducible direct summands of the state cone the state lies in.
Call a positive map $T : A \rightarrow A$ {\em nondisturbing} on
state $\omega$ if $T(\omega) = c_{\omega} \omega$ for some nonnegative
constant $c_{\omega}$ that in principle could depend on the state.
Say such a map is {\em nondisturbing} if it is nondisturbing on all
pure states.\footnote{ Of course,  
%if we condition on information
%obtained, 
this definition permits mixed states to be disturbed by a
nondisturbing map---that can be viewed as something like an
inevitable ``epistemic'' disturbance associated with obtaining
information.}
%% The important thing is that
%% for pure states, conditioning on the fact that a nondisturbing $T$
%% occured is dividing by $c_\omega$, and the state is unchanged.
%%
%% We could also say a {\em norm-nonincreasing} map
%% $T$ is nondisturbing if there is some {\em instrument}, that is, set of
%% norm-nonincreasing maps $T_i$, summing to a norm-preserving
%% one, such that $T$ is one of the $T_i$, and $\sum_i T_i = \id$.  That
%% is, a
A norm-nonincreasing map nondisturbing in this sense is precisely
the type of map that can appear associated with some measurement
outcome in an operation that, averaged over measurement outcomes,
leaves the state (pure or not) unchanged.
%Thus
%% it is
%% only the information associated with which of a set of nondisturbing
%% maps has occured, that can be obtained without (on average)
%% disturbing the state.  The fact that the mixed state is disturbed conditional
%% on the measurement result may be considered ``purely epistemic'' because if we forget the
%% information, we have the original state.

A cone $C$ in a vector space $V$ is a {\em direct sum} of cones $D$
and $E$ if $D$ and $E$ span disjoint (except for $0$) subspaces of
$V$, and every element of $C$ is a positive combination of vectors
in $D$ and $E$.  A cone is irreducible if it is not a nontrivial
direct sum of cones.  Every finite-dimensional cone is uniquely
expressible as a direct sum $C = \oplus_i C_i$ of irreducible cones
$C_i$.  Information about which of the summands a state is in should
be thought of as ``inherently classical'' information about the
state.
%% , analogous to information about an
%% Abelian observable (``charge'') associated with a superselection rule in
%% quantum mechanics (which is a special case).  A classical system (simplex)
%% is a direct sum of one-dimensional cones.

\begin{theorem}
The nondisturbing maps on a cone that is a sum $C = \oplus_i C_i$ of
irreducible $C_i$, are precisely the maps $M = \sum_i c_i \id_i$,
where $\id_i$ is the identity operator on the summand $V_i$ and the
zero operator elsewhere, and $c_i$ are arbitrary nonnegative
constants.
\end{theorem}

For a nondisturbing map, $c_\omega$ can depend only on the
irreducible component a state is in; thus, the fact that a
nondisturbing map has occured can give us no information about the
state within an irreducible component. In other words, as claimed,
only inherently classical information is contained in the fact that
a nondisturbing map has occured.
%% Also,
%% within an irreducible component, even a {\em mixed} state must be
%% undisturbed by learning that such a map has occurred: it
%% turns out that the ``purely epistemic'' disturbance to mixed states we
%% countenanced in our definition of nondisturbance, can only
%% concern the relative weights of the irreducible components, and is
%% disallowed within those components.

The existence of information that cannot be obtained without
disturbance is often taken to be the principle underlying the
possibility of quantum key distribution, so the fact that it is
generic in nonclassical theories in the framework leads us (with
Barrett and Leifer) to conjecture that secure key distribution,
given an authenticated public channel, is possible in all
nonclassical models.

\section{Composite systems and Entanglement}

Given two physical systems, represented by abstract state spaces $A$
and $B$, we naturally want to describe {\em composite} systems
having $A$ and $B$ as subsystems. We make the (non-trivial)
assumption that a bipartite state $\omega$ on systems $A$ and $B$ is
defined by a joint probability weight
\[\omega : [0,u_A] \times [0,u_B] \rightarrow {\Bbb R}.\] In other words, we assume
that the joint state of two systems be determined by the
probabilities assigned {\em local} measurements, i.e., measurements
pertaining to the two systems separately.
Barrett \cite{Barrett:2005a} calls this the {\em global state
assumption};  we follow \cite{BBLW08} in calling it the 
{\em local observability} condition.\footnote{The condition is violated by 
both real and quaternionic quantum
mechanics \cite{Klay:1987}.}  Such a bipartite state is {\em non-signaling}
iff, for all observables $E$ on $A_1$,
\[\omega_{E}(a,b) : = \sum_{a \in E}
\omega(a,b)\] is independent of $E$, and similarly for the other
component.  One can show (see \cite{Klay:1987,Wilce92a}) that $\omega$ is
non-signaling iff it extends to a bilinear form on $A_{1}^{\ast}
\times A_{2}^{\ast}$.  It is clear that, conversely,
any bilinear form $\omega$ that is {\em positive} in the sense
that $\omega(a,b) \geq 0$ for all $(a,b) \in A^{\ast}_{+} \times
B^{\ast}_{+}$, and normalized by $\omega(u_A, u_B)$, defines a
state.  Thus, we can identify the set of possible bipartite states
with the space $B(A^{\ast}, B^{\ast})$ of bilinear forms on
$A^{\ast} \times B^{\ast}$, ordered by the cone of positive forms.

For our purposes, it will be convenient to identify the space
$B(A^{\ast}, B^{\ast})$ with the tensor product $A \otimes B$,
interpreting the pure tensor $\alpha \otimes \beta$ of states
$\alpha \in A, \beta \in B$ as the form given by $(\alpha \otimes
\beta)(f,g) = f(\alpha)g(\beta)$ where $f \in A^{\ast}, g \in
B^{\ast}$. We call a form $\omega \in A \otimes B$ {\em positive}
iff $\omega(a,b) \geq 0$ for all $(a,b) \in A^{\ast}_{+} \times
B^{\ast}_{+}$. If $\omega$ is positive and $\omega(u_A, u_B) = 1$,
then $\omega(a,b)$ can be interpreted as a joint probability for
effects $a \in A^{\ast}$ and $b \in B^{\ast}$.
%Conversely, one
%can show (see [4] and [10]) that any assignment of joint
%probabilities consistent with a no-signaling requirement must be
%bilinear.
Thus, the most general model of a composite of $A$ and $B$
consistent with our definition of a joint state and the no-signaling
requirement,  is the space $A \otimes B$, ordered by the cone of all
positive forms, and with order unit given by $u_{A} \otimes u_{B} :
\omega \mapsto \omega(u_{A}, u_{B})$. This gives us an abstract
state space, which we term the {\em maximal tensor product} of $A$
and $B$, and denote $A \maxtensor B$. At the other extreme, we might
wish to allow only product states $\alpha \otimes \beta$, and
mixtures of these, to count as bipartite (normalized) states. This
gives us the {\em minimal tensor product}, $A \mintensor B$. These
coincide if either factor is classical, that is, if $\Omega_{A}$
or $\Omega_{B}$ is a simplex \cite{Namioka:1969}; in general,
however, the maximal tensor product allows many more states than the
minimal. A state in $\Omega_{A \maxtensor B}$ not belonging to
$\Omega_{A \mintensor B}$ is {\em entangled}.

More generally, we define a {\em composite} of $A$ and $B$ to be
{\em any} state space $AB$ consisting of bilinear forms on $A^{\ast}
\times B^{\ast}$, ordered by a cone $AB_{+}$ of positive forms
containing every product state $\alpha \otimes \beta$, where $\alpha
\in \Omega_{A}$ and $\beta \in \Omega_{B}$.  Equivalently, $AB$ is
a composite iff $A \mintensor B \leq AB \leq A \maxtensor B$ (where,
for abstract state spaces $A$ and $B$, $A \leq B$ means that $A$ is
a subspace of $B$, that $A_{+} \subseteq B_{+}$, and that $u_{A}$ is
the restriction of $u_{B}$ to $A$). More generally still, a
composite of $n$ state spaces $A_1,...,A_n$ is a state space $A$ of
$n$-linear forms on $A_1^{\ast} \times \cdots \times A_{n}^{\ast}$,
ordered by any cone of positive forms containing all product states.
In such a composite, define the {\em conditional state space} of a 
subset $J$ of the parts as the set of states obtainable by conditioning
on all product effects of parts not in $J$.  
Following \cite{BBLW08} we call such a composite 
{\em regular} if all of its conditional state spaces are also composites.

We'll call the formalism outlined above, in which the unnormalized
effects on a system are the full dual cone of the state cone,
composites are non-signalling, contain all product states, and satisfy
local observability, simply ``the framework'' in order to avoid
repetition of this list of assumptions.  It is the framework used, for
the most part, in \cite{BBLW2007}, \cite{BBLW08}, and \cite{BDLT2007a}, 
which are the sources of most of the results we outline below.
Barrett's framework \cite{Barrett:2005a} is essentially the same
except it does not require the effects to be the full dual cone.

\section{Cloning, Broadcasting, Bit Commitment and Teleportation}

Many of the most celebrated results of quantum information theory
turn out to have much more general formulations in terms of abstract
state spaces. An example, first pointed out by Barrett \cite{Barrett:2005a}, is
that the impossibility of universal cloning, far from being a
specifically quantum phenomenon, is {\em generically} non-classical,
in the sense that it is a feature of {\em all} probabilistic
theories involving non-classical state spaces.  In this section, we
survey recent work in this direction.

\subsection{No-cloning and no-broadcasting}

Let $AA$ be any composite of $A$ with (a copy of) itself, and let
$\phi : A \rightarrow AA$ be any positive, norm-preserving mapping.
We say that $\phi$ {\em clones} a set $\Gamma$ of (normalized)
states iff $\phi(\alpha) = \alpha \otimes \alpha$ for all $\alpha
\in \Gamma$, and that $\phi$ {\em broadcasts} $\Gamma$ iff
$\phi(\alpha)_{A} = \phi(\alpha)_{B} = \alpha$ for all $\alpha \in
\Gamma$.  In \cite{BBLW2007, Barnum:2006}, it is shown that (i) $\Gamma$ is
broadcastable iff it is contained in the convex hull of a set of clonable
states, and (ii) $\Gamma$ is clonable iff jointly distinguishable.
The standard quantum no-broadcasting theorem is an easy corollary
(and, indeed, this provides the easiest known proof of the latter).
%[explain].
In fact, \cite{Barnum:2006} shows slightly more: for {\em any} 
positive map, the set of states it broadcasts is {\em precisely} such 
a simplex generated by distinguishable states.

\subsection{Bit commitment}

Quantum theory has mixed states whose representation as a convex
combination of pure states is not unique.  So do {\em all}
nonclassical theories: uniqueness of the decomposition of mixed
states into pure states is an easy characterization---sometimes used
as a definition---of simplices (see, for example, the proof in
\cite{Barrett:2005a}).  While we are not aware of any quantum information
processing task whose possibility is directly traced to the
non-unique decomposability of mixed states into pure, this was
certainly proposed as a possible basis for quantum bit commitment
schemes, though (as shown in \cite{Bennett84a} for their proposed
scheme, and in \cite{Mayers97a,Lo97a} for more elaborate schemes)
these schemes do not work because of entanglement.

Bit commitment is an important cryptographic primitive in which one
party (``Alice'') can perform an act that commits her, vis-a-vis a
partner (``Bob'') to the value of a bit in such a way that she is able, at
will, to reveal the committed bit value to Bob and have him accept it
(perhaps on the basis of some tests he performs) as genuine.  The
protocol is binding: once committed to a bit value, Alice will not be
able to get Bob to accept the other value as the revealed bit.  It is
also hiding: once Alice has committed, Bob knows she has commited, but
knows nothing about the value of the bit.  Information-theoretically
secure bit commitment (applicable even to parties with unlimited
computational power) is impossible in classical probability theory.
In \cite{BDLT2007a} it is shown that the existence of bit commitment
protocols is universal in nonclassical theories in the framework, provided that the
tensor products used do not permit entanglement. Consider theories
generated by a finite set $\Sigma$ of ``elementary'' systems modeled
by convex sets $\Omega, \Gamma...$ in finite dimensions, containing at
least one nonsimplex.  Let it be closed under the minimal, or
separable, tensor product, which we write with the ordinary tensor
product symbol $\otimes$.

\noindent {\em The protocol}. Let a system have a non-simplicial,
convex, compact state space $\Omega$ of dimension $d$, embedded as
the base of a cone of unnormalized states in a vector space $V$ of
dimension $d+1$.
%% \begin{Def} A point $\mu$ in a compact convex set
%% $\Omega \subset \R^d$ is
%% said to be {\em exposed} if there is an affine hyperplane $H_\mu$
%% whose intersection with $\Omega$ is precisely $\{\mu\}$.
%% \end{Def}
The  protocol uses a state $\mu$ that has two distinct
decompositions into finite disjoint sets $\{\mu^0_i\}, \{\mu^1_j\}$
of exposed states,  that is, \beqa \label{2Decomp} \omega = \sum_{i
= 1}^{N_0} p_i^0 \mu^0_i = \sum_{j=1}^{N_1} p_j^1 \mu^1_j, \eeqa

A state $\mu_i^b$ is {\em exposed} if there is a measurement outcome
$a_i^b$ that has probability $1$ when, and only when, the state is
$\mu_i^b$; the protocol exists for all nonclassical systems because,
as we show, any non-simplicial convex set of affine dimension $d$
always has a state $\omega$ with two decompositions (as above), into
{\em disjoint} set of states whose total number $N_0 + N_1$ is $d+1$
The disjointness and the bound on cardinality are used in the proof
of exponential security given in \cite{BDLT2007a}.

In the honest protocol, Alice first decides on a bit $b \in \{0,1\}$
to commit to.  She then draws $n$ samples from $p^b$, obtaining a
string $\ve{x} = (x_1,x_2,\ldots,x_n )$.  To commit, she sends the
state $\mu^b_{\ve{x}} = \mu^b_{x_1} \otimes \mu^b_{x_2} \otimes \ldots
\otimes \mu^b_{x_n}$ to Bob.  To reveal the bit, she sends $b$ and
$\ve{x}$ to Bob.  Bob measures each subsystem of the state he has.  On
the $k$-th subsystem, he performs a measurement, (which will depend on
$b$) containing the distinguishing effect for $\mu^b_{x_k}$ and
rejects if the result is not the distinguishing effect.  If he obtains
the appropriate distinguishing effect for every system, he
accepts. The protocol is perfectly sound (if Alice and Bob are honest, Bob
never accuses her of cheating and always obtains the correct bit),
perfectly hiding (if Alice is honest, Bob cannot gain any information
about the bit until Alice reveals it), and has an exponentially low
probability, in $n$, of Alice's successfully cheating (is exponentially
binding).

\subsection{Conditioning and teleportation protocols}

%[Here, let's omit discussion of regularity, focussing on the easier
%situation where the composite is $A \mintensor BC$.]

In the standard quantum teleportation protocol \cite{BennettTele},
Alice and Bob share an entangled state; Alice possesses an ancillary
system in an unknown state. By making a suitable entangled
measurement on her total system, and instructing Bob to make
suitable unitary corrections on his wing of the shared system, Alice
can guarantee that the final state of Bob's wing is identical to the
unknown initial state of her ancilla (which, in compliance with the
no-cloning theorem, is irrevocably altered).

In recent work \cite{BBLW08} with Jonathan Barrett and Matthew Leifer,
we consider what such a protocol looks like in the setting of abstract
state spaces. We find that teleportation is possible in a much broader
class of probability theories than just quantum and classical
theory. However, unlike the no-cloning and no-broadcasting theorems,
which are {\em generically} non-classical (at least in finite
dimensions), the existence of a teleportation protocol imposes a real
constraint on nonclassical theories.

If $AB$ is a composite of state spaces $A$ and $B$, we can define
for any normalized state $\omega \in AB_{+}$ and any effect $a \in
A$, both a {\em marginal} state $\omega_{A}( - ) = \omega(-, u_{B})$
and a {\em conditional} state $\omega^B_{a}$ defined by the condition 
$\omega^B_a(b) =
\omega(a,b)/\omega_{A}(a)$ (with the usual proviso that if
$\omega_{A}(a) = 0$, the conditional state is also $0$). We shall
also refer to the partially evaluated state $\omega_{B}(a) :=
\omega( a, - )$ as an {\em un-normalized} conditional state.

Note that any state $\omega \in AB$ gives rise to a positive
operator $\hat{\omega} : A^{\ast} \rightarrow B$, given by
$\hat{\omega}(a)(b) = \omega(a,b)$.  That is, $\hat{\omega}(a)$
is the ``un-normalized" conditional state corresponding to 
conditioning on $a$. As a partial converse, any
positive operator $\psi : A^{\ast} \rightarrow B$ with $\psi(u_{A})
\in \Omega_{B}$---that is, with $\psi^{\ast}(u_{B}) := u_{B} \circ
\psi = u_{A}$---corresponds to a state in the maximal tensor
product $A \maxtensor B$. Dually, any effect $f \in (AB)^{\ast}$
yields an operator $\hat{f} : A \rightarrow B^{\ast}$, given by
$\hat{f}(\alpha)(\beta) = f(\alpha \otimes \beta)$; and any positive
operator $\phi : A \rightarrow B^{\ast}$ with $\phi(\alpha) \leq
u_{B}$ for all $\alpha \in \Omega_{A}$---that is, with $\|\phi\|
\leq 1$---corresponds to an effect in $(A \mintensor B)^{\ast}$.

Suppose now that $f$ is an effect in $(A \mintensor B)^{\ast}$ and
$\omega$ is a state in $B \maxtensor C$. Then, for any $\alpha \in
A$, it is not difficult to check that
\begin{equation}(\alpha\otimes \omega)^{C}_{f} = \hat{\omega}
(\hat{f}(\alpha))\\ \|\hat{\omega} \circ
\hat{f}(\alpha)\|.\end{equation} Notice that, in consequence, if $c$
is any effect in $C^{\ast}$, $f \otimes c$ is positive on product
states of the form $\alpha \otimes \omega$, with $\alpha \in A$ and
$\omega \in B \maxtensor C$, and hence, defines an effect in $A
\mintensor (B \maxtensor C)$.

Put differently, this shows that $(A^{\ast} \maxtensor B^{\ast})
\mintensor C^{\ast}$ is an ordered subspace of $(A \mintensor (B
\maxtensor C))^{\ast}$. This allows us to interpret equation
(\theequation) as follows: if the tripartite system $ABC = A
\mintensor (B \maxtensor C)$  is in a state $\alpha \otimes \omega$,
with $\alpha$ unknown, then conditional on securing measurement
outcome $f$ in a measurement on $A \maxtensor B$, the state of $C$
is, up to normalization, a {\em known function of} $\alpha$.
%We call this {\em remote evaluation.}
This is very like a teleportation
protocol. Indeed, suppose that $C$ is a copy of $A$, and that $\eta
: A \rightarrow C$ is a specified isomorphism allowing us to match
up states in the former with those in the latter:

\begin{definition} With notation as above, $(f,\omega)$ is a (one-outcome, post-selected) {\em teleportation
protocol}
iff there exists a positive, norm-contractive {\em correction map}
$\tau : C \rightarrow C$ such that, for all $\alpha \in A$,
$\tau(\alpha \otimes \omega)^{C}_{f} = \eta(\alpha)$.\footnote{One
could also allow protocols in which the correction has a nonzero
probability to fail. For details, see \cite{BBLW08}.}
\end{definition}

If $(f,\omega)$ is a teleportation protocol, the un-normalized
conditional state of $\alpha \otimes \omega$ is exactly
$\hat{\omega}(\hat{f}(\alpha))$. If we let $\mu := \hat{\omega}
\circ \hat{f}$, the normalized conditional state can be written as
$\hat{\mu}(\alpha)/u(\hat{\mu}(\alpha))$. Thus, $(f,\omega)$ is a
teleportation protocol iff there exists a norm-contractive mapping
$\tau$ with $(\tau \circ \mu)(\alpha) = \|\mu(\alpha)\|\eta$ for all
$\alpha \in \Omega_{A}$.  $\|\mu(\alpha)\| > 0$ (which is the probability
of getting measurement outcome $f \otimes u_C$ on state $\alpha \otimes \omega$)
is the probability that the teleportation protocol succeeds on state $\alpha$. 
%$some positive constant $s$.
%[details: one can show $s$ independent of $\alpha$....---state
%theorem!]

%\begin{theorem} With notation as above, $(f,\omega)$ is a teleportation protocol iff $\mu : = \hat{\omega} \circ \hat{f}$ is proportional
%to an isomorphism $(A,u_{A}) \simeq (C,u_{C})$; in this case, $\tau
%: (C,u_{C}) \simeq (C,u_{C})$ is also an isomorphism. \end{theorem}

Henceforth, we simply identify $C$ with $A$, suppressing $\eta$.
We say a composite $ABA$ {\em supports a conclusive teleportation protocol} 
if such a protocol $(f,\omega)$ exists with $f,\omega$ allowed states and
effects of the composite.    
Note that if $(f,\omega)$ is a teleportation protocol on a regular
composite [{\em AW: ``regular composite" not yet defined...?}] $A B A$ of $A$, $B$ and (a copy of) $A$, then, as $f$
lives in $(AB)^{\ast} \leq (A \mintensor B)^{\ast}$ and $\omega$
lives in $BA \leq B \maxtensor C$, one can also regard $(f,\omega)$
as a teleportation protocol on $A \mintensor (B \maxtensor A)$.

\begin{theorem}[\cite{BBLW08}] $A \mintensor (B \maxtensor A)$ supports a conclusive teleportation protocol iff $A$ is order-isomorphic to the
range of a compression (a positive idempotent mapping) $P :
B^* \rightarrow B^*$. \end{theorem}

\begin{corollary} If $B$ is order-isomorphic to $A^*$ then 
$A \mintensor (B \maxtensor A)$ supports a conclusive
teleportation protocol. \end{corollary}

\begin{corollary} If $A$ can be teleported through a copy of
itself, then $A$ is weakly self-dual. \end{corollary}

In order to {\em deterministically} teleport an unknown state
$\alpha \in A$ through $B$, we need not just one entangled effect
$f$, but an entire observable's worth.

\begin{definition} A {\em deterministic teleportation protocol} for $A$ through $B$ consists of
an observable $E = (f_1,...,f_n)$ on $A \otimes B$ and a state
$\omega$ in $B \otimes A$, such that for all $i = 1,...,n$, the
operator $\hat{f_i} \circ \hat{\omega}$ is physically
invertible.\end{definition}

By {\em physically invertible}, we mean its inverse is a norm-contractive
positive map.  

The following result provides a sufficient condition (satisfied,
e.g., by any state space $A$ with $\Omega_{A}$ a regular polygon)
for such a protocol to exist.

\begin{theorem}[\cite{BBLW08}] Let $A = B$. Suppose that $G$ is a finite group acting transitively on the pure states of $A$,
and let $\omega$ be a state such that $\hat{\omega}$ is a
$G$-equivariant isomorphism. For all $g \in G$, let $f_{g} \in (A
\maxtensor A)^{\ast}$ correspond to the operator
\[\hat{f}_{g} = \frac{1}{|G|} \hat{\omega}^{-1} \circ g.\]
Then $E = \{f_{g} | g \in G\}$ is an observable, and $(E, \omega)$
is a deterministic teleportation protocol.\end{theorem}

\section{Categorical considerations}

In this section we briefly sketch one way of relating the
above-described convex framework to the category-theoretic framework
for theories developed by Abramsky and Coecke and by Selinger.  If an
abstract state space and its dual ``effect space" provide an abstract
probabilistic model, one would like to say that a probabilistic {\em
  theory} is a class of such models, perhaps closed under appropriate
operations producing models of composite systems from models of their
components.  Each such model should be equipped with a dynamical
semigroup of allowed evolutions, which must be positive maps.  A
natural way of formalizing such an approach is to say that a theory is
a category whose objects are abstract state spaces and whose morphisms
are positive linear maps between these spaces, composed as usual.

Such a category will in addition have a ``unit object'', referred to
as $I$, consisting of $\R$ understood as a one-dimensional vector
space over itself, ordered by its usual (and in fact unique) positive
cone $\R_+$, with order unit equal to the identity function from $\R$
to $\R$.  Since the morphisms from, say, $A$ to $B$ live in the real
vector space $\cl(A,B)$ of linear maps from $A$ to $B$, we can and
will require $\Hom(A,B)$ to itself be a (pointed, generating) positive
cone.  We will require that $\Hom(I,A)$ be isomorphic to $A_+$.  The
isomorphism $\eta: A_+ \rightarrow \Hom(I,A); \omega \mapsto f_\omega$
is taken to satisfy $f_\omega(1)=\omega$.  That is, the states of $A$
can be viewed as morphisms from the unit to $A$.  Since
(not-necessarily normalized) effects are positive maps from $A$ to
$I$, all effects are potentially elements of $\Hom(I,A)$.  We will
require $\Hom(I,A)$ to contain $u_A$ in its interior; thus, the
order-unit structure for each object specifies a distinguished
morphism $u_A \in \interior \Hom(A,I)$.  This formalism easily accomodates
one small but important divergence from the framework described above:
by requiring all Hom-sets to be positive cones, but not necessarily
requiring every positive linear functional to belong to $\Hom(A,I)$,
we have not enforced the assumption we made above, that the cone of
unnormalized effects was the full dual cone of the cone of states.
(Rather, our formalism implies only that it is a subcone of the dual
cone.)  We'll call an object for which $\Hom(A,I)$ {\em is} the cone dual to $\Hom(I,A)$
{\em saturated}.  
We will call a theory in which all objects are saturated {\em
locally saturated}.  The ability to formulate theories that are not
locally saturated is crucial for dealing with, for example, convex
versions of Rob Spekkens' theory of toy bits \cite{Spekkens:2004}.
Relaxing the assumption of local saturation likely has nontrivial 
implications for informatic phenomena.

The interpretation of such a category is somewhat loose, but we may
think of the morphisms as {\em processes} that a system may undergo.
This is essentially the Abramsky-Coecke point of view on categorially
formulated theories.  Since morphisms may connect objects with
different objects, this allows a process to change the nature of a
system, to one described by a different object.  Whether this is an
actual physical change in the nature of a system, or whether we want
to regard it as a change in point of view, is a somewhat delicate
point of interpretation.  It could represent, for example, discarding
or disregarding part of a system, or cobbling up a new system and
combining it with the old one, or receiving a new system delivered
from some other agent.  We'll remain fairly agnostic on this point, as
various interpretations may be useful for various applications.  In
our setting, the {\em norm-contractive} morphisms are the
operationally relevant ones, corresponding to something that can
actually be realized in the system being modelled by the formalism;
the other ones are merely mathematically convenient to include in the
formalism.

In particular, the morphisms $f_\omega: I \rightarrow A$ may be viewed
as ``bringing up a new system, $A$, prepared in state $\omega$"'.  The
morphisms $A \rightarrow I$ are effects.  Let $g$ be such an effect;
$g(\omega)$, the probability of observing effect $g$ when $A$ is
prepared in state $\omega$, may be represented as $g(f_\omega(1))$, or
if one prefers, $g \circ f_\omega (1)$.  More generally, for any 
state $\omega$ of  $A$, and process $\phi: A \rightarrow B$, the probability
of the state undergoing the process is calculated by applying the order unit, 
i.e. as $u_B \circ \phi \circ f_\omega(1)$.  (Strictly speaking, these
are not guaranteed to be {\em probabilities} unless all morphisms involved
are norm-contractive.)
The structure of a category does several useful things for us.  Since
morphisms are positive, and the composition of morphisms must be a
morphism, it enforces that we will never get a negative number out of
any chain of morphisms $I \rightarrow I$.  In other words, no process
constructed (in any way) from other processes will ever have negative
probability. Similarly, processes composed out of norm-contractive
processes will have probabilities bounded above by $1$.

The interpretation also explains the meaning of our assumption that
all Hom-sets are cones: operationally, it means that for any two
processes, there is another process consisting of doing one or the
other of the first two proceses, conditional on the outcome of a
``coin-flip'' (where the coin may be chosen to have arbitary bias).

We call such a category of positive maps {\em saturated} if there is
no way to enlarge it by adding positive maps to some $\Hom(A,B)$
(while keeping all the morphisms we started with).
Note that a category may be saturated without being locally saturated,
and vice versa.  
We call a category {\em locally \Hom-saturated at
$(A,B)$} if the subcategory whose objects are $A,B,I$ and whose
$\Hom$-sets are those of the original category, is saturated, and 
{\em locally $\Hom$-saturated} if it is $\Hom$-saturated at every pair of
objects.  The category whose objects are finite-dimensional mixed
quantum state spaces and whose morphisms are completely positive maps,
for example, is locally saturated, but neither locally
$\Hom$-saturated nor saturated.    There are examples
of categories that are locally saturated, $\Hom$-saturated, and saturated:
for example, the categories generated by taking a fixed set of objects 
and closing under coupling the state spaces $\Hom(I,A)$ 
by the minimal tensor product, and determining
the remaining $\Hom$-sets by local saturation and local $\Hom$-saturation.
A similar construction, but closing under the maximal tensor product, 
also exhibits all three properties.  
     
%% [{\em AW: Why is it saturated? If I add a non-CP map $f : A \rightarrow B$ to $\Hom(A,B)$ for some pair of quantum systems, 
%% and then add all maps arising from composition of $f$ with existing maps, the resulting maps are still positive, so 
%% surely I get a new category? To be sure, we lose the monoidal structure (the tensor product no Mlonger makes sense), but isn't that another story?}]

Various kinds of structure may be added to our categories.  Here, we
consider structures representing one object's being a composite of
other objects.  In \cite{BBLW08} we introduced a very general notion
of multipartite composite, though not in an explicitly
category-theoretic framework.  We now briefly consider a specific way
of modeling compositeness in a category of positive maps, with a view
to bringing out connections with the work of Abramsky and Coecke and
of Selinger.  Namely, we consider such categories that are, in
addition, {\em monoidal}, with monoidal tensor written $\otimes$.  Our
first order of business is to compare $A \otimes B$ with the notion of
composite introduced above.  The bifunctoriality of $\otimes$ implies
that $\Hom(A \otimes B, C \otimes D)$ contains $\Hom(A,C) \times
\Hom(B,D)$ where elements $(\alpha, \beta)$ of this Cartesian product
are written $\alpha \otimes \beta$, and turn out to be just the usual
tensored pairs of linear maps.  This implies that the composite state
space includes all product states, and the composite effect space
$\Hom(A \otimes B, I)$ contains all product effects.  Hence the states
of the composite must be positive on all product effects, so we have
two of the properties we required of a bipartite composite in the
framework of earlier sections.  More precisely, we have them for
saturated objects, hence for all objects in a locally saturated
theory; the statement in the previous sentence is the natural
generalization of the two requirements to theories whose
objects are not necessarily saturated.

However, the third condition on composites in the above framework,
that of {\em local observability}, does not necessarily hold in such a
category.  Local observability, though extremely natural and still
permitting an extraordinarily wide range of composite systems and of
informatic phenomena, is an assumption with relatively substantive
implications.  For example, it makes bit commitment significantly
harder, by ruling out ``intrinsically nonlocal'' degrees of freedom.
Allowing the latter permits theories \cite{BS08a} which effectively
have bipartite ``lockbox-key pairs'' along the lines suggested by John
Smolin \cite{Smolin2003a} in which a bit may be put by Alice and
``stranded'' between Alice and Bob when she sends Bob one of the
systems as her commitment.

As remarked above for the case of quantum systems, the category, which
we'll call ${\bf C^*CPOS}$, of $C^*$-algebra state spaces and
completely positive maps, is not saturated.  We can enlarge it by
adding maps $\varphi: A \rightarrow B$ that are positive but not
completely positive, as long as we do not add the maps $\varphi \otimes
\id: A \otimes C \rightarrow B \otimes C$ except in the trivial case
$C=I$.\footnote{One can add {\em all} the positive maps if one
  likes, obtaining a category we'll call ${\bf C^*POS}$.} In the
category ${\bf C^*CPOS}$, one can introduce a natural symmetric
monoidal structure, which in the case of quantum systems just gives us
the standard tensor product.  If we attempt to add positive but not
completely positive maps while maintaining monoidality, we fail, for
we are forced to include all maps $f \otimes \id$, and these can
fail to be positive.  We call a {\em monoidal} category of positive
maps saturated if one cannot enlarge it by adding positive maps, while
maintaining monoidality; we'll usually say it's {\em monoidally
  saturated} to avoid confusion in cases, such as ${\bf C^*CPOS}$,
which are not saturated as categories, but are saturated as monoidal
categories.

In the case where local observability happens to hold, the
associativity and bifunctoriality of $\otimes$ together imply that
multipartite composites are {\em regular} in the sense of
\cite{BBLW08} and the preceeding section.  It should not be hard to
generalize the notion of regularity to systems lacking local
observability, and we expect the appropriate generalization to hold
for all monoidal categories of positive maps as defined above.  There
are very natural monoidal categories of abstract state spaces and positive maps
for which local observability fails.  The category whose objects are
spaces of real symmetric $n \times n$ matrices ordered by their
positive semidefinite cones and with order unit given by the trace,
and whose morphisms are completely positive maps between such spaces,
is a case in point.  This theory is, of course, just the
finite-dimensional mixed-state version of real (as opposed to complex)
quantum mechanics.  Here if $A_+$ is the unnormalized density matrices
(i.e. the positive semidefinite matrices) on real $n$-dimensional
Hilbert space $H_A$, and $B$ is the same for $H_B$, then $A \otimes B$
is the vector space of real symmetric matrices on $H_A \otimes H_B$,
which is higher dimensional than the tensor product of $A$ and $B$ as
vector spaces.  Nevertheless, even in this case, the projection onto
the tensor product of the underlying vector spaces is a composite in
the sense of our earlier framework.  We suspect that using this
composite as $A \otimes B$ would still give a symmetric monoidal
category.  However, in this case there is a compelling reason for
going beyond it: to preserve the representation of the state space as
the set of density matrices on
a real Hilbert space.  This has nice properties like self-duality and
homogeneity of the state space; preserving such properties is a
more abstract motivation, probably applicable to this case, for
sometimes going beyond local observability.

\section{Conclusion}

We have reviewed some of our recent work, with various sets of
collaborators, on information processing and informatic phenomena in
the convex operational framework for theories, and outlined a way of
connecting it to the category-theoretic framework.  The
information-processing results fall into two basic classes.  

First, there are results which elucidate the distinction between
classical and nonclassical information, thereby elucidating the
distinction between classical and nonclassical theories, as we can
characterize classical theories as ones in which {\em all} information
is classical.  From these results, two types of classical information
emerge within general theories in the framework.  First, there is {\em
  intrinsically} classical information.  Emerging as the condition for
information to be extractable from a system without disturbing it,
this is information that tells us which summand in a direct-sum
decomposition of the cone of states the state is in.  It is
intrinsically classical in the sense that when a cone has a direct sum
decomposition, by definition every extremal ray is in one summand or
the other; this is to say that no pure state of the theory is a {\em
  superposition} of, and no mixed state exhibits {\em coherence}
between, states in different summands.  The notion of superposition in
the convex framework was investigated in \cite{Beltrametti93a,
  Lahti85, LahtiBugajski85}, and we will explore it and our
mixed-state extension of the notion to that of {\em coherence}, in
more depth elsewhere.  Second, there is a notion of information that
might be called {\em facultatively classical}, corresponding to a {\em
  classical substructure} in a model.  This is information about which
of a set of perfectly distinguishable states of the theory the system are in.
Intrinsically classical information is of course also facultatively
classical, but not vice versa: the theory may allow coherence between
perfectly distinguishable states, and then distinguishing the states
will disturb other states, including some pure ones, that have
coherence between the distinguishable states.  But the information
about which state we have within a facultatively classical set of
states---a simplex within the convex set of states, whose vertices are
perfectly distinguishable states---{\em can} be gathered without
disturbing {\em those states} (i.e., the vertices will not be
disturbed), and similarly this kind of information can be broadcast.

Second, there are results that establish informatic properties of
broad classes of theories, but do not just provide a way of
demarcating classical from non-classical theories.  These include very
general and simple results linking to well-known properties of
composites such as entanglement (which generalizes easily to this
framework), such as the result that nonclassical theories without
entanglement support exponentially secure bit commitment.  They also
include results that bring out the importance of new properties of
theories, or at least ones that have been less stressed in the quantum
literature, such as weak self-duality of state spaces, and the
existence of states and maps that implement that self-duality, which is
sufficient for teleportation.  Our study of teleportation, besides
yielding interesting necessary and sufficient conditions for
conclusive teleportation in a three-part composite, has also turned up
nice sufficient conditions for {\em deterministic teleportation},
relating the existence of a high degree of symmetry of the state
space---a transitive group of automorphisms---and, again, appropriate
bipartite states---to teleportation.  We expect that further
investigation of which classes of theories permit bit commitment, will
bear the same sort of conceptual fruit.

Finally, we related this convex framework to the categorial framework
in which much recent work has explored quantum theory and
protocols---and increasingly, non-quantum foil theories as well.  As
we have implemented it, the framework treats systems that may be built
up from multiple subsystems, or at any rate analyzable in terms of
subsystem structure---using notions of {\em composite} or {\em regular
  composite} that are somewhat looser, structurally, than the
symmetric monoidal structure usually used in the category-theoretic
approach.  But formulating convex theories as categories of positive
maps, as we do in the last section, is very natural and provides a
natural way to relax two assumptions that are usually present in our
use of the convex framework but which are substantive and which, as we
have noted, it will be sometimes desirable to relax.  These are that
for each system the effect cone is the full dual of the state cone (``local
saturation''),  and that the
probabilities of product effects determine the state (``local
observability'').  It also shows how the convex approach can provide a
wealth of concrete examples of theories that may or may not have a
monoidal structure, but that are in either case much more general than
the strongly (a.k.a. dagger) compact closed ones that have formed the
main object of study in the categorial approach.  Using these concrete
examples, the extensive results that exist concerning ordered linear
spaces, their face lattices, and their symmetries, and the currently
developing, and potentially very rich, theory of how this order
structure behaves in composite systems, especially how it interacts
with monoidality, we believe the convex operational approach and the
categorial approach will continue to fruitfully interact in the
project of characterizing quantum theory by its informatic properties
and its information-processing powers, not just in contradistinction
to classical theory, but in contrast to the panoply of other theories
that both approaches provide as foils.  A promising avenue along which
progress will likely be made is to understand the implications of
strong compact closure in the convex setting, and the implications of
natural properties of ordered linear spaces, notably homogeneity and
self-duality, for categorially structured theories.  Both strong
compact closure and homogeneous self-duality single out significantly
restricted classes of theories, classes that include quantum theory.  We
venture to guess that these classes may be somewhat similar, and that
combining concepts of categorial and of convex origin may help us,
both in narrowing down the class of theories and thus obtaining formal
results characterizing quantum theory within a broad space of
glistering foils set off to th'quantum world, and in understanding the
operational content, and implications for information processing, of
the principles used in such characterization results.  This will give
us a fascinating, exciting, and useful perspective on the essence of
quantum theory, as the unique theory, within a broad framework for
theories, in which information, and the ways in which it can be
processed, has a specific set of fundamental properties.  We expect
this understanding to have pragmatic applications in the development
of new quantum information processing protocols, and in understanding
the limits on such protocols.  But we also expect it may contribute to
progress on the vexing questions of how quantum theory is to be
interpreted, what it implies for the way in which our physics relates
to the world, and how it fits with the rest of physical theory.

\bibliographystyle{plain}
%\bibliography{barnummasterjan17_08c}
%use c when I find it; probably on laptop
%%%\bibliography{barnummasterjan17_08}

\end{document}